\documentstyle[prb,twocolumn,aps,psfig,epsf]{revtex}
\begin{document}
\draft

\twocolumn[\hsize\textwidth\columnwidth\hsize\csname @twocolumnfalse\endcsname

\title{Numerical Study of the Incommensurate Phase in Spin-Peierls
Systems}

\author{A. E. Feiguin, J. A. Riera, A. Dobry and H. A. Ceccatto}

\address{Instituto de F\'{\i}sica Rosario, Boulevard 27 de Febrero 210
bis, 2000 Rosario, Argentina}
\maketitle

\begin{abstract}
We analyze several properties of the lattice solitons
in the incommensurate phase
of spin-Peierls systems using exact diagonalization and quantum Monte 
Carlo. These systems are modelled by an antiferromagnetic Heisenberg
chain with nearest and next-nearest neighbor interactions coupled to 
the lattice in the adiabatic approximation. Several relations among 
features of the solitons and magnetic properties of the system have 
been determined and compared with analytical predictions. We have 
studied in particular the relation between the soliton width and the
spin-Peierls gap. Although this relation has the form predicted
by bosonized field theories, we have found some important 
quantitative differences which could be relevant to describe 
experimental studies of spin-Peierls materials. 
\end{abstract}


\vskip2pc
]
\narrowtext

\section{Introduction}
\label{intro}

One-dimensional or quasi-one-dimensional magnetic systems show many
fascinating properties which continue to attract an intense
theoretical activity. One of these properties is the presence of
a spin gap in antiferromagnetic Heisenberg chains with integer 
spin\cite{haldane} and in ladders.\cite{ladders} 
Another, particularly complex, system which presents a spin gap
is the spin-Peierls (SP) system. In this system a Heisenberg chain
coupled to the lattice presents an instability at a critical
temperature, $T_{SP}$, below which a dimerized lattice pattern 
appears and a spin gap opens in the excitation spectrum.\cite{pytte}

The interest in the spin-Peierls phenomena was recently revived after
the first inorganic SP compound, CuGeO$_3$, was found.\cite{hase}
This inorganic material allows the preparation of better samples than
the organic SP compounds and hence several experimental techniques
can be applied to characterize the properties of this
system.\cite{regnault}
Besides, this compound can be easily doped with magnetic and 
non-magnetic impurities, leading to a better understanding of its
ground state and excitations.\cite{lussier}

Spin-Peierls systems present also a very rich and interesting 
behavior in the presence of an external magnetic field.
Below the spin-Peierls transition temperature,
and for magnetic fields $H$ smaller than a critical value
$H_{cr}(T)$, the system is in its spin-Peierls phase, characterized
by a gapped nonmagnetic ($S^z=0$) ground state with a 
dimerized pattern or alternating nearest-neighbor (NN) interactions. 
For $T < T_{tc} < T_{SP}$, at $H=H_{cr}(T)$ a transition occurs 
from the dimerized phase to a gapless incommensurate (IC) state 
characterized by a finite magnetization, $S^z > 0$. $T_{tc}$ is 
the temperature of the point at which the dimerized, incommensurate 
and uniform phases meet.
The dimerized-IC transition was predicted by some 
theories\cite{cross2} to be of first order at low temperatures, and 
this is the behavior found in experimental studies\cite{hase2,loosd}.
Other theories predict that this transition is a second order 
one.\cite{fujita}

A simple picture of the dimerized-IC transition can be obtained by
mapping the Heisenberg spin chain to a spinless fermion system
by a Jordan-Wigner transformation.
The effect of the magnetic field favoring a nonzero $S^z$ due to the
Zeeman energy can be interpreted as a change in the band filling
of the equivalent spinless fermion system. As a result, the momentum
of the lattice distortion moves away from $\pi$ as
$\tilde{q} = (1-S^z/N) \pi$, where $N$ is the number of sites on the
chain. However, since {\it umklapp} processes pin the momentum at 
$\pi$ up to a critical field $H_{cr}(T)$, the lattice distortion will 
remain a simple dimerization and the magnetic ground state will
remain a singlet.\cite{crossfisher}
Theoretical\cite{fujita,nakano,buzdin} and 
experimental\cite{kiry,fagot} studies indicate that the lattice 
distortion pattern in the IC phase corresponds to an array of 
solitons.
A complementary picture indicating how a soliton lattice could appear 
as a consequence of the finite magnetization in the IC phase is the 
following. Let's assume that the dominant contribution to the 
magnetic ground state comes from a state of NN singlets or dimers.
An up spin replacing a down spin destroys a singlet and gives rise to 
two domain-walls or solitons separating regions of dimerized order 
which are shifted in one lattice spacing with respect to each other. 
Each soliton carries a spin-1/2. Due to the spin-lattice coupling it
is expected that the lattice solitons are driven by these magnetic 
solitons.

The soliton formation in spin-Peierls systems has been studied 
analytically by bosonization techniques applied to the spinless
fermion model.\cite{affleck}
The coupling to the lattice is treated usually in
the adiabatic approximation. The resulting field-theory formalism
has lead to important results, the most remarkable being the 
relation between the soliton width and the spin-Peierls gap,
$\xi \sim \Delta^{-1}$.\cite{nakano}
Although this formalism has been extended to a Heisenberg model
with competing NN and next-nearest-neighbor (NNN) antiferromagnetic
interactions\cite{zang2,dobryriera}, it presents some unsatisfactory 
features.

In the first place, there are some recent experimental 
results\cite{kiry} for the soliton width in the IC phase in CuGeO$_3$ 
indicating a disagreement with the theoretical prediction. Although 
there might be a contribution to the soliton width coming from 
magnetic\cite{zang2} or elastic\cite{dobryriera} interchain 
couplings which would explain at least partially this
disagreement, it is also possible that the differences could be due
to several approximations involved in the bosonized field theory.
One should take into account that these theories are valid in 
principle in the long wave-length limit, and the applicability of
their results to real materials can not be internally assessed.
Then, our first motivation to start a numerical study of the IC phase
in spin-Peierls systems is to measure the importance of
these approximations in the analytical approach.

In the second place, the field theory approach does not provide a 
detailed dependence of the magnitudes involved in terms of the 
original parameters of the microscopical models. For example, even 
for the simplest case\cite{nakano} the expression obtained for the 
spin-wave velocity must be replaced by the exact one known from 
Bethe's exact solution of the Heisenberg chain. In this sense, 
numerical studies could give information about how the relevant 
magnitudes depend on the original parameters without further 
approximations.

With these motivations, in this article we want to initiate   
the study of the incommensurate phase in SP systems using numerical 
methods. These methods give essentially exact results for finite
clusters, and they can be used to 
check various approximations required by the analytical approaches 
and the validity of their predictions. Besides, the
numerical simulations provide a detailed information of the 
dominant magnetic and lattice states. In Section \ref{lanczos} we
present the model considered and we study several features of the 
soliton
formation in the IC phase using the Lanczos algorithm. In particular
we analyze the effect of NNN interactions on the soliton width.
In Section \ref{monte} we perform Monte Carlo simulations using the
world line algorithm --which allows us to study larger chains than
the ones accessible to the Lanczos algorithm-- in order to reduce
finite size effects.

\section{Exact diagonalization study}
\label{lanczos}

The one-dimensional model which contains both the antiferromagnetic
Heisenberg interactions
and the coupling to the lattice is:

\begin{eqnarray}
{\cal H} &=& J \sum_{i = 1}^N (1 + (u_{i+1}- u_{i}))\;
{\bf S}_i \cdot {\bf S}_{i+1} \nonumber \\
&+& J_2 \sum_{i = 1}^N {\bf S}_i \cdot {\bf S}_{i+2} 
+ \frac{K}{2} \sum_{i=1}^N (u_{i+1}- u_{i})^2
\label{hamtot}
\end{eqnarray}
\noindent 
where ${\bf S}_i$ are the spin-1/2 operators and $u_i$ is the
displacement of magnetic ion $i$ with respect to its
equilibrium position. Periodic boundary conditions are imposed.
The first term, which corresponds to the nearest neighbor (NN)
interactions, contains the spin-lattice coupling in the
adiabatic approximation.
The second term contains the AF NNN interactions, which were proposed
in Refs. [\onlinecite{rieradobry,castilla}] to fit the experimental
magnetic susceptibility data in CuGeO$_3$. Several other properties 
of this material have been reasonably described using this 
model.\cite{haas,rierakoval,poilblanc} As in 
Ref. [\onlinecite{rieradobry}], we assume for simplicity that the
lattice distortion does not affect the second neighbor interactions.
In principle, the NNN interactions should be corrected by a term
proportional to $(u_{i+2}- u_{i})$ which vanishes in the 
dimerized phase but not necessarily in the incommensurate phase.
This correction should be important precisely in the region around
a soliton. It is customary to introduce the frustration constant
$\alpha= J_2/J$. The estimated value of $\alpha$ in CuGeO$_3$
varies between 0.24 (Ref. [\onlinecite{castilla}]) and 0.36
(Ref. [\onlinecite{rieradobry}]). In this second case, $\alpha$
is larger than the critical value $\alpha_c \approx 0.2411$ above
which in the absence of dimerization a gap opens in the excitation
spectrum.\cite{okamoto}

Our purpose is to study numerically Hamiltonian (\ref{hamtot})
with exact diagonalization (Lanczos) techniques and by Monte Carlo 
simulations. In this latter case, in order to avoid the well-known
sign problem due to the frustration, we will consider only 
the diagonal second neighbor interaction

\begin{eqnarray}
{\cal H}_2^{zz} =
J_2^{z} \sum_{i = 1}^N S_i^{z} S_{i+2}^{z},
\label{h2n-zz}
\end{eqnarray}
\noindent 

instead of the isotropic NNN interactions (second term of Eq.
\ref{hamtot}).

It is quite apparent that the main numerical
difficulty is related to the handling of the set of
displacements $\{u_i\}$, which in principle can take arbitrary
values to describe the various distortion patterns present
in the dimerized and IC phases of the system. 
These displacements are calculated self-consistently by the
following iterative procedure. First, we introduce the bond
distortions defined as $\delta_i = (u_{i+1}- u_{i})$. Then,
the equilibrium conditions for the phononic degrees of freedom:

\begin{eqnarray}
\frac{\partial \langle {\cal H} \rangle}{\partial \delta_i} + 
\lambda = 0
\label{eqcond}
\end{eqnarray}
\noindent
lead to the set of equations:
\begin{eqnarray}
J \langle {\bf S}_i \cdot {\bf S}_{i+1} \rangle + K \delta_i -
{J \over N} \sum_{i=1}^N \langle {\bf S}_i \cdot {\bf S}_{i+1} 
\rangle = 0 ,
\label{distor-eqn}
\end{eqnarray}
\noindent
which satifies the constraint $\sum_i \delta_i =0$.
This constraint has been included in Eq. (\ref{eqcond}) through
the corresponding Lagrange multiplier $\lambda$.
The expectation values are taken with respect to
the ground state of the system. 
The iterative procedure starts with an initial
distortion pattern $\{ \delta_i^{(0)} \}$, which in general
we choose at random. At the step $n$, with a distortion 
pattern $\{ \delta_i^{(n-1)} \}$, we diagonalize
Hamiltonian (\ref{hamtot}) using the Lanczos algorithm and
compute the correlations 
$\langle {\bf S}_i \cdot {\bf S}_{i+1} \rangle$. We replace
these correlations in Eq. (\ref{distor-eqn}) and the new set
$\{ \delta_i^{(n)} \}$ is obtained. We repeat this iteration
until convergence. Essentially the same procedure is followed
in the quantum Monte Carlo algorithm, as it is discussed in
Section \ref{monte}.

We have applied this exact diagonalization procedure to determine 
the distortion patterns in the 20 site chain at $T=0$. 
In the first place we consider the case of $S^z=0$. 
As mentioned above, this corresponds to a dimerized lattice, 
i.e. $\delta_i = (-1)^i \delta_0$.
Notice that for this simple case, the equilibrium distortion 
amplitude $\delta_0$ could be determined in an 
easier way by computing the energies of the spin part of Hamiltonian
for a set of values of $\delta_0$. Then, adding the elastic energy
and interpolating one obtains the minimum total energy.
We have performed this calculation in order to check our iterative
algorithm.

The results for $\delta_0$ vs. $K$, for $S^z=0$, are
shown in Fig.1 for 
$\alpha = 0.0$, 0.2 and 0.4, and $J_2^{z} = 0.2$, and 0.4.
It can be seen that, as expected, for $\alpha >0$ the dimerized 
state
is more favorable and this leads to a larger $\delta_0$ for a given
$K$. To a lesser extent this trend is also present for $J_2^{z} > 0$.

The dependence of $\delta_0$ with $K$ can be inferred from
the scaling relation between the energy and the dimerization,
$E_0(\delta_0)-E_0(0) \sim \delta_0^{2\nu}$ (plus logarithmic 
corrections) with $\nu=2/3$, in principle valid for
$\alpha < \alpha_c$ and small 
$\delta_0$.\cite{crossfisher,spronken,laukamp}
Then, it is easy to obtain $\delta_0 \sim K^{-3/2}$,
a relation which is approximately satisfied by our numerical 
data. The fact that $\delta_0$ vanishes at a finite value 
$\hat{K}$ of the elastic constant, is just a finite size 
effect. By diagonalizing chains of $N=12$, 16 and 20 sites, 
for $\alpha=0$, we have verified that $\hat{K}$ increases with the 
lattice size, as it can be seen in Fig. \ref{fig1.5}, and it 
should eventually diverge in the bulk limit.

\vspace{-1.5cm}
\begin{figure}
\psfig{figure=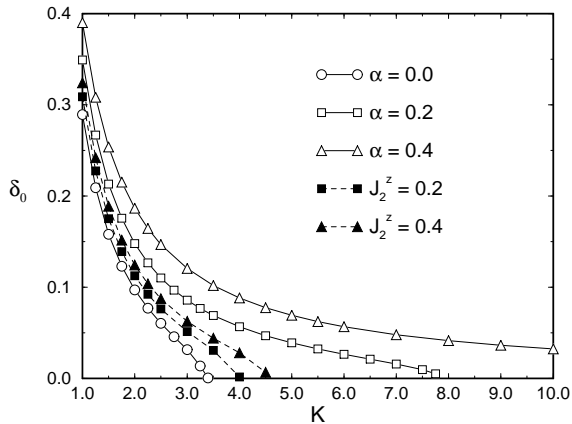,width=9.00cm,angle=-90}
\vspace{0.1cm}
\caption{Dimerization amplitude vs. elastic constant 
obtained by exact diagonalization in the 20 site chain,
$S^z=0$, for various values of $\alpha = J_2/J$ and $J_2^{z}$.}
\label{fig1}
\end{figure}

Once we have determined the equilibrium distortion as a function 
of $K$, we are able to compute the singlet-triplet spin gap, defined 
as the following difference of ground state energies:
\begin{eqnarray}
\Delta = E_{0,dim}(S^z=1)-E_0(S^z=0)
\label{spingap}
\end{eqnarray}
\noindent
It is worth to emphasize that $E_{0,dim}$ is the ground state energy
of the system for $S^z=1$ with the dimerization obtained for $S^z=0$
and the same set of parameters.
The results of this calculation are shown in Fig. \ref{fig2}.
Consistently with the larger $\delta_0$ shown in Fig. \ref{fig1}, 
the gap increases with $\alpha$. The effect of $J_2^{z}$ is
much weaker than that of the isotropic second neighbor interaction
which is not surprising since the 1D ground state magnetic structure,
with a dominant dimerized state,
has essentially a quantum (off-diagonal) origin.
This small increase in $\Delta$ for a given $K$ is consistent with
the small increase in $\delta_0$ shown in Fig. \ref{fig1}.
The corresponding scaling relation, $\Delta \sim K^{-1}$, obtained 
from the relation between the singlet-triplet gap and
the dimerization, $\Delta \sim \delta_0^{2/3}$,
is again reasonably satisfied by our numerical data.

\vspace{-1.cm}
\begin{figure}
\psfig{figure=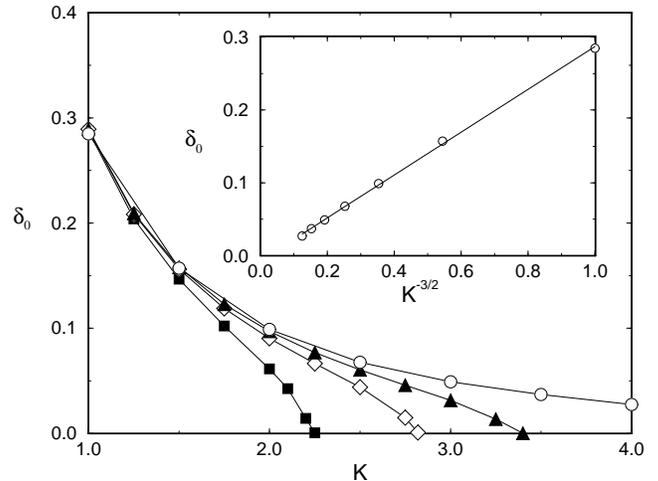,width=9.00cm,angle=-90}
\vspace{0.3cm}
\caption{Dimerization amplitude vs. elastic constant 
obtained by exact diagonalization for $N=12,16,20$ (solid
squares, diamond and triangles, respectively) and Monte Carlo
simulations for $N=64$ (open dots), with
$\alpha = 0$. The inset shows the expected scaling behavior
$\delta_0 \sim K^{-3/2}$ for $N=64$. }
\label{fig1.5}
\end{figure}

\vspace{-1.cm}
\begin{figure}
\psfig{figure=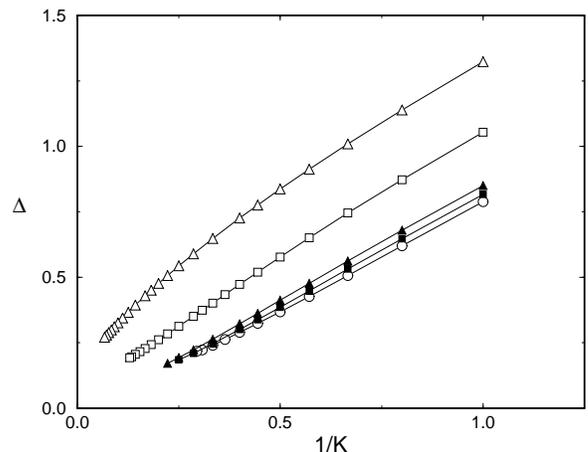,width=9.00cm,angle=-90}
\vspace{0.1cm}
\caption{Singlet-triplet gap vs. elastic constant 
obtained by exact diagonalization in the 20 site chain,
for various values of $\alpha$ and $J_2^{z}$. The symbols
have the same meaning as in Fig. \ref{fig1}}.
\label{fig2}
\end{figure}


\vspace{-1.5cm}
\begin{figure}
\psfig{figure=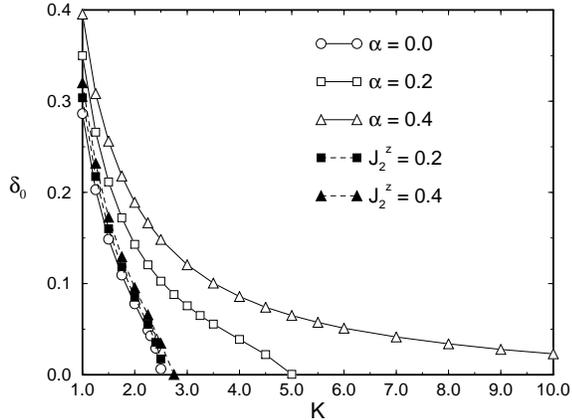,width=9.00cm,angle=-90}
\vspace{0.1cm}
\caption{Dimerization amplitude vs. elastic constant 
obtained by exact diagonalization in the 20 site chain, 
$S^z=1$, for various values of $\alpha$ and $J_2^{z}$.}
\label{fig3}
\end{figure}

We now consider the case of $S^z=1$, which corresponds to the
incommensurate region just above the dimerized-incommensurate
transition. We have determined the distortion pattern for
a 20 site chain using the iterative procedure described above. 
As discussed
at the beginning of this section, the two solitons or domain 
walls separating dimerized regions are clearly distinguishable.
(A typical pattern can be seen in Fig. \ref{soliton}.)
The maximum distortion $\delta_0$, shown in Fig. \ref{fig3}, 
presents similar behavior as the one shown in Fig. \ref{fig1}
corresponding to $S^z=0$. In particular, the fact that $\delta_0$
vanishes at a finite $K$ is again due to finite size effects.

In order to compute the soliton width, we use the following
form to fit the  numerically obtained distortion patterns:

\begin{eqnarray}
\delta_i = (-1)^i \tilde{\delta} \tanh \left(\frac{i-i_0 -
{\frac d 2}}{\xi}
\right)\tanh \left(\frac{i-i_0+{\frac d 2}}{\xi}\right),
\label{fittanh}
\end{eqnarray}
\noindent
which corresponds to
modeling each soliton as an hyperbolic tangent, as obtained in the
analytical approach to this problem.\cite{nakano} 
The amplitude $\tilde{\delta}$, the soliton width $\xi $, and the
soliton-antisoliton distance $d$, 
are the parameters determined by the numerical fitting. 
The amplitude $\tilde{\delta}$ should be equal to the maximum 
distortion $\delta_0$ defined above for well separated solitons,
i.e. $d \gg \xi$.
The main limitation of this calculation arises in the region where,
for a given $\alpha$, $K$ is so large that
the solitons have a substantial overlap in the 20 site chain, and 
the fitting function (\ref{fittanh}) is no longer appropriate. 
In this case, the elliptic sine should be used to describe the 
soliton lattice. This is the region where finite size effects are 
important, as it was discussed above with
respect to Figs. \ref{fig1} and \ref{fig3}.
However, this situation is not directly relevant to experiment
since in real materials the solitons are well-separated.\cite{kiry}

\begin{figure}
\psfig{figure=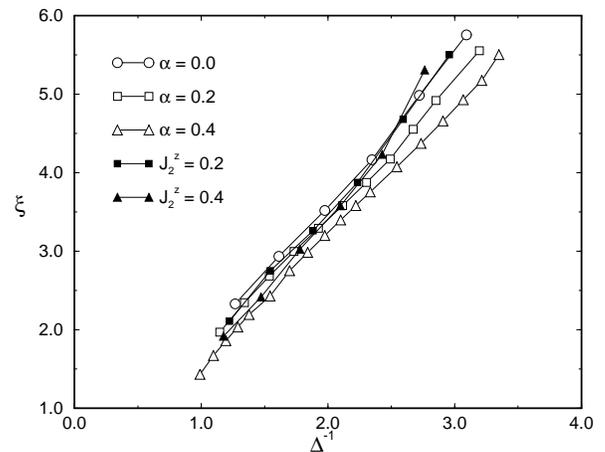,width=9.00cm,angle=-90}
\vspace{0.1cm}
\caption{Soliton width vs. singlet-triplet spin gap 
obtained by exact diagonalization in the 20 site chain,
$S^z=0$, for various values of $\alpha$ and $J_2^{z}$.}
\label{fig4}
\end{figure}

We show in Fig. \ref{fig4} the soliton width as a function of the gap 
$\Delta$ for the 20 site chain, for the same values of $\alpha$
and $J_2^{z}$ as before.
It can be seen that the there is a linear dependence of the soliton
width with the inverse of the gap. This behavior is consistent with
the theoretical prediction:\cite{nakano}

\begin{eqnarray}
\xi =v_s/\Delta,
\label{width-gap}
\end{eqnarray}
where $v_s$ is the spin-wave velocity for $\alpha < \alpha_c$.
It was recently shown that the relation (\ref{width-gap}),
originally obtained for the unfrustrated chain,\cite{nakano} is also 
valid in the presence of frustration.\cite{dobryriera} For 
$\alpha > \alpha_c$, 
$\Delta$ contains a contribution from the frustration due to the 
presence of a gap even in the absence of dimerization.

A linear fitting of these curves in the region $\xi > 2.5$ gives the
slopes 1.87, 1.70 and 1.63 , for 
$\alpha =0.0,\;0.2$ and $0.4$ respectively.
Recently, a numerical study\cite{fledder} has proposed the law:
$v_s=\frac \pi 2 (1-1.12 \alpha )$ in the bulk limit
for $\alpha < \alpha_c$,
From this law one gets $v_s =$ 1.57, 1.22, for 
$\alpha =0.0$ and $0.2$ respectively. We can observe that the slopes
obtained by fitting the curves shown in Fig. (\ref{fig4})
are systematically larger than these values of $v_s$. 
Besides, the effect of $\alpha$ is weaker in the numerical
data than that predicted by Eq. (\ref{width-gap}). 
For $\alpha= 0.4 > \alpha_c \approx 0.2411$, we have estimated 
$v_s$ by fitting the excitation dispersion relation
$\varepsilon(k) = E_{0,dim}(S^z=1,k) - E_0(S^z=0,k=0)$ with the law
$\varepsilon(k)^2 = \Delta^2 + v_s^2 k^2 + c k^4$ around $k=0$
and $\delta_i =0$.
For $L=20$ we obtained $v_s = 0.707$, a value which is also
smaller than
the slope of the curve $\xi$ vs. $1/\Delta$ for $\alpha=0.4$ in
Fig. (\ref{fig4}).
This disagreement between the prediction obtained by the 
continuum bosonized theory and the numerical results could be
due to the approximations involved in the former or to finite
size effects present in the latter. The study of much larger
lattices than those considered in this section will be done
in the following section using quantum Monte Carlo simulations.
On the other hand, for the case of $J_2^z = 0.4$ the slope is
actually {\it larger} ($\approx 2.1$) than the value obtained
for the Heisenberg
chain with NN interactions only. This effect is opposite to that
of the isotropic NNN interactions and it will be further discussed
in the next section.

\vspace{-1.cm}

\section{Monte Carlo simulations}
\label{monte}

In order to treat longer chains than those considered in the Lanczos
diagonalization study of the previous section, we have implemented a
world-line Monte Carlo algorithm\cite{WLMC} suited to this problem. 
The partition function is re-expressed as a functional integral over
wordline configurations, where the contribution on each imaginary-time
slice is given by the product of the two-site evolution matrix 
elements,

\begin{eqnarray}
W_{i,i+1}(\tau )=\langle S_{i,\tau }^zS_{i+1,\tau }^z\left| {\rm e}^{-
\Delta \tau J_i {\bf S}_i \cdot {\bf S}_{i+1}}\right| S_{i,\tau
+\Delta \tau }^zS_{i+1,\tau +\Delta \tau }^z\rangle 
\nonumber
\end{eqnarray}
\noindent
where $J_i = J(1 + \delta_i)$.
These matrix elements are the Boltzmann weights associated with a bond
($ i,i+1$) in a time step $\Delta \tau =1/mT$ in the Trotter direction,
where $ T $ is the temperature and $m$ is the Trotter number. Since the
exchange couplings depend on the lattice displacements, these matrix
elements are site dependent.

\begin{figure}
\begin{center}
\leavevmode
\psfig{figure=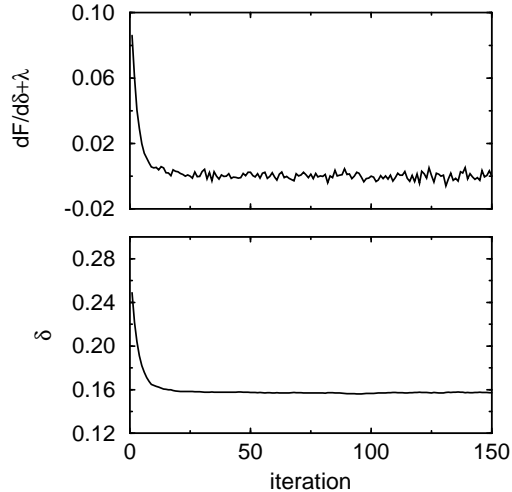,width=9.0cm,angle=-90}
\end{center}
\caption{Minimization of the free energy for 
a uniform dimerized chain. We plot the derivative
of the free energy and the parameter $\delta$ along
the successive iterations.}
\label{150}
\end{figure}

We implemented the algorithm with the addition of a dynamic 
minimization of the free energy with respect to the lattice 
displacements. Starting from a given initial configuration (random
distribution of spins and a dimerized pattern for the lattice 
displacements) we typically considered $2\times 10^3$ sweeps for 
thermalization. During the next $4\times 10^3$ sweeps we measured the 
derivative of the magnetic free energy, which, in the limit of
$T \rightarrow 0$, is given by

\begin{equation}
\frac{\partial {\cal F}_M}{\partial \delta _i}=J\langle \langle 
{\bf S}_i {\bf \cdot S}_{i+1}\rangle \rangle _T \ .
\label{DF}
\end{equation}
Leaving 3 sweeps between each measurement for de-correlation this 
produces 
$10^3$ independent values to obtain the thermal average. With this
free-energy gradient we corrected the displacements according to 
(\ref{distor-eqn}) and repeated the procedure, including the 
$2\times 10^3$ sweeps for thermalization since the spins have to 
accommodate to the new lattice distorsions. Once the displacement
pattern is stabilized within statistical fluctuations ---we
typically considered $\sim $150 iterations, see Fig. \ref{150}--- we 
performed
measurements of several quantities. For this we obtained 100 independent
groups of $10^3$ measurements each, following the same procedure as
described above, {\it i.e.}, i) thermalization, ii) measurements of 
$\frac{\partial {\cal F}_M}{\partial \delta _i}$
and observables, and iii) correction of the displacement 
pattern due to statistical fluctuations.

In our calculations we considered chains of 64 sites with periodic 
boundary conditions and a temperature $T=0.05J$. We checked that this 
value is low enough to study ground-state properties by comparison 
with measurements at even lower temperatures. On the other hand, at 
higher temperatures the soliton is not observed and there is no 
definite pattern of lattice displacements. We took $m=80$ for the 
Trotter number, which is large enough to reproduce the Lanczos 
results on smaller chains (see Fig. \ref{fig1.5}). 
For some particular quantities like the energy gap, which require 
more precision, we considered also $m=160.$ In addition, comparison 
with results for a longer chain with $N=128$ indicates that in the 
parameter range of our calculations the Monte Carlo
results have no sizeable finite-size effects.

\vspace{-1.cm}

In Fig. \ref{fig1.5} we show the Monte Carlo results for the 
homogeneous dimerization of the 64 site chain in the $S^z=0$ 
subspace as a function of the elastic constant $K$, together 
with the Lanczos results for smaller chains. Notice that in
the parameter range considered the 64 site chain does not have 
the finite size effects present for smaller chains, namely, the 
vanishing of $\delta_0$ for finite values of $K$. The inset shows the
expected scaling behavior $\delta \propto K^{-3/2}$ discussed in the
previous section. As a further check, we have also reproduced the 
scaling behavior of the energy gain $E_0(\delta_0)-E_0(0)$ 
and gap with $\delta_0$ with a measured exponent $\nu =2/3$
within statistical errors.

\vspace{-1.cm}
\begin{figure}
\begin{center}
\leavevmode
\epsfxsize=9.00cm
\epsffile{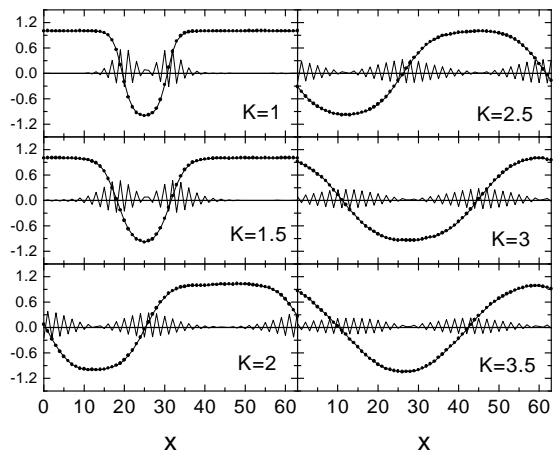}
\end{center}
\caption{Lattice distortion patterns and local magnetizations of the
64 site chain obtained
by Monte Carlo, for $S^z=1$ and different values of $K$. In every 
panel the maximum lattice distortion is normalized to one, so that
they cannot be directly compared.}
\label{soliton}
\end{figure}

\vspace{-1.cm}
\begin{figure}
\begin{center}
\leavevmode
\psfig{figure=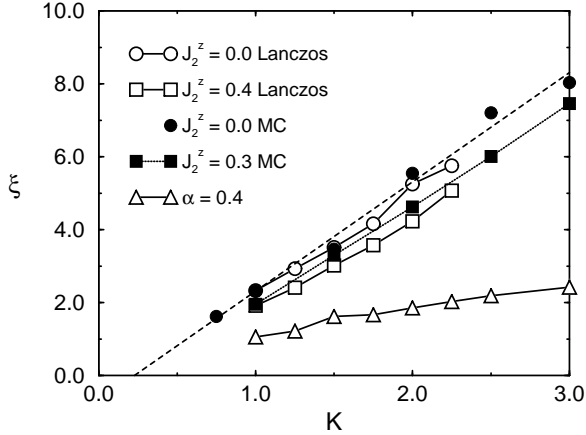,width=9.00cm,angle=-90}
\end{center}
\caption{Soliton width vs. elastic constant $K$ obtained by Monte
Carlo simulations in the 64 site chain for $J_2^{z}=0.0$ and 0.3, 
together with Lanczos results for the 20 site chain, $J_2^{z}=0.0$
and 0.4, and $\alpha = 0.4$. The dashed line corresponds
to a linear fit to the Monte Carlo results for $J_2^{z}=0$.}
\label{xi-k}
\end{figure}

The soliton structure in the subspace with 
$S^z=1$ is given in Fig. \ref{soliton}, where we plot the 
displacement envelope 
$\widetilde{\delta }_i=(-)^i\delta_i$ and the local magnetization 
$\langle S_i^z\rangle $ , for different values of the elastic 
constant $K.$ Notice that the displacements are normalized by
their maximum values (shown in Fig. \ref{fig1.5}) and the local 
magnetization by the classical value $S=1/2.$ 
Consequently, the size of lattice distortions in different panels 
cannot be directly compared. For small values of $K$ there is a well
defined soliton-antisoliton structure in the distortion pattern, with
the associated local magnetization following a staggered order.
There is a net $1/2$ spin density near each domain wall,
which makes the excess $S^z=1$. As in the previous section, we 
fitted a two-soliton solution (\ref{fittanh}),
with $\tilde{\delta }_0=1$ because of the normalization adopted. The 
results for the soliton width $\xi$ are shown in Fig. \ref{xi-k}. 
For increasing values of $K$ the soliton width grows until the
displacement profile resembles a sine law (see Fig. \ref{soliton}).  
This sinusoidal pattern is typical of the 
soliton lattice, observed for large values of $S^z$.
It can be seen that the scaling $\xi \sim K$ obtained in 
[\onlinecite{nakano}] is well reproduced in the whole parameter range
considered, as indicated by the linear fit to the data (dashed line). 
This figure shows that the soliton width for $J_2^z = 0.3$ also 
presents a linear dependence with $K$. These features observed in
the 64 site chain are qualitatively similar to those present
in the 20 site chain as determined by exact diagonalization.
Besides, it can seen in this figure that the reduction of $\xi$ is
much stronger when the isotropic NNN is taking into account.

We have performed a simple study on the soliton-antisoliton 
interaction. For this study we fixed the distortion pattern to the 
law (\ref{fittanh}) with the previously fitted value of $\xi ,$ and
considered increasing values of $d.$ For small $K$ ($\leq 2J$) we 
found that the total energy becomes a constant (within statistical 
fluctuations) when $ d\geq 4\xi ,$ which implies that the 
soliton-antisoliton pairs shown in the left panels of Fig. 4 are not 
interacting. This was confirmed by allowing the lattice distortion 
to evolve starting from a pattern like (\ref{fittanh}) with an 
initial separation larger than $d,$ which produces the same result 
for $\xi $ and the total energy.  

\vspace{-1.cm}
\begin{figure}
\begin{center}
\leavevmode
\psfig{figure=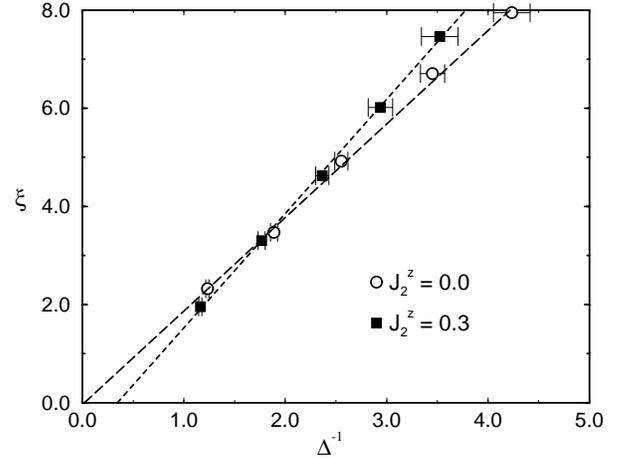,width=9.00cm,angle=-90}
\end{center}
\caption{Soliton width vs. inverse of singlet-triplet gap obtained 
by Monte Carlo simulations for $J_2^{z}=0.0$ and 0.3. The horizontal 
error bars give the estimated error in the determination of the
gap $\Delta$.}
\label{xi-gap}
\end{figure}

Next, we study the behavior of the soliton width $\xi $ with the 
spin-Peierls gap $\Delta $. That is, we compare the quantity 
$\xi $ that characterizes the $S^z=1$ soliton state, with the
singlet-triplet excitation gap $\Delta $ above the dimerized $S^z=0$
ground state. As shown in Fig. \ref{xi-gap}, these two quantities are 
inversely related to each other, as discussed in the previous 
section. The slope of the linear fit is $1.9$, very close to the 
value 1.87 obtained by exact diagonalization of the 20 site chain in  
the previous section.  This result confirms the disagreement between
the numerical results with the analytical prediction pointed out in
Section \ref{lanczos}. Also shown in Fig. \ref{xi-gap} are 
the results
for $J_2^z = 0.3$. A linear fit to these results leads to a slope
$\approx 2.3$ , i.e. larger than the value corresponding to
$J_2^z = 0.0$. This increase of the slope between $\xi$ and
$\Delta^{-1}$ is consistent with the result obtained for the
20 site lattice by exact diagonalization and $J_2^z = 0.4$.
This behavior should be contrasted with the {\it reduction}
of the slope found for the isotropic NNN interaction.
A possible explanation of this behavior could be the following.
As discussed in the previous section, the term ${\cal H}_2^{zz}$
leads to a smaller increase of the spin gap than the fully isotropic
NNN interaction. On the other hand, the Ising interaction could be
more effective in punishing the excess $\langle S^z \rangle$
which appears around a soliton leading to a smaller reduction 
of the soliton width than the one caused by the isotropic
term, as it can be seen in Fig. \ref{xi-k}.
A more detailed study of the Hamiltonian in the presence of the
term of ${\cal H}_2^{zz}$ is clearly necessary to fully understand
this behavior.

\begin{figure}
\psfig{figure=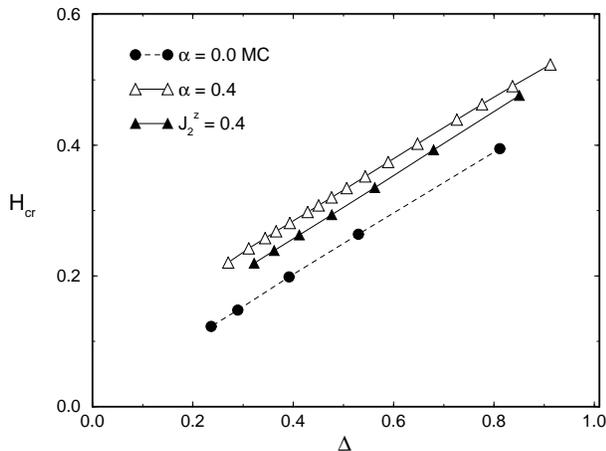,width=9.00cm,angle=-90}
\vspace{0.1cm}
\caption{$H_{cr}$ vs. spin gap obtained by Monte Carlo in the 
64 site chain for $\alpha =J_2^{z}=0.0$ and by exact diagonalization
in the 20 site chain, for $\alpha = J_2^{z} = 0.4$.}
\label{hcritvsgap}
\end{figure}

Finally, it is possible to estimate the critical value of the 
magnetic field at zero temperature. By adding a Zeeman term to the
Hamiltonian (\ref{hamtot}), $-g \mu_B S^z H$ ($\mu_B$: Bohr's 
magneton), $H_{cr}$ may be calculated as:
\begin{eqnarray}
H_{cr} = E_0(S^z=1)-E_0(S^z=0)
\label{hmagcrit}
\end{eqnarray}
in units of $g \mu_B$. $E_0(S^z=1)$ is the ground state energy of 
(\ref{hamtot}), and then $H_{cr} < \Delta$, which is the value
expected of a gapped system in the absence of magneto-elastic
coupling.
The behavior of $H_{cr}$ as a function of $\Delta$ is shown in Fig. 
\ref{hcritvsgap} for the 64 site chain, $\alpha=J_2^z=0.0$, 
and for the 20 site chain, $\alpha=J_2^z=0.4$.
It is apparent a linear dependence over all the range
studied, which is in agreement with the mean-field 
prediction\cite{pytte,crossfisher}, $H_{cr} \approx 0.84 \Delta$.
However, we obtain a coefficient considerable smaller,
$H_{cr}/\Delta \approx 0.47$, almost independent of $\alpha$.
This value is also smaller than twice the soliton formation energy
calculated in Ref. [\onlinecite{nakano}]. The finite value at the
origin of the
curves corresponding to $\alpha=J_2^z=0.4$ is a finite size effect.

\vspace{-1.cm}

\section{Conclusions}
\label{conclu}

In this article we have analyzed the magnetic soliton lattice in 
the incommensurate phase of spin-Peierls systems using numerical
methods. There is a remarkable agreement between the results obtained
by exact diagonalization using the Lanczos algorithm and those
obtained by quantum Monte Carlo with the world-line algorithm. The 
relations among various
features of the solitons and magnetic properties of the system have 
been determined and compared with analytical results.
Our starting point is a microscopical model proposed to describe
several properties of CuGeO$_3$, consisting of a 1D AF Heisenberg
model with nearest and next-nearest neighbor interactions. 

In the
first place we have not detected any crossover in the behavior of the
quantities examined as $\alpha$, the ratio of NNN to NN interactions,
becomes greater than $\alpha_c$ at least for the small chains
considered. That is, there are only smooth changes as $\alpha$ varies
between 0.0 and 0.4. The most important effect of the competing
NNN interaction is a {\it reduction} of the soliton width $\xi$ as a 
function of the inverse of the singlet-triplet spin gap $\Delta$.
Furthermore, the effect of the diagonal term (\ref{h2n-zz}) is much
less important and in some cases even qualitatively different
to that of the isotropic NNN term.

Although several functional forms predicted by continuum analytical
theories have been confirmed by our numerical data, there are some
important quantitative differences. The most important disagreement
between our numerical results and the analytical predictions is
related to the coefficient in the relation $\xi \sim \Delta^{-1}$,
i.e. we have obtained a systematically higher value than the
theoretical value which is the spin-wave velocity. The estimated value
of $H_{cr}/\Delta$ is also noticeable smaller than the mean-field 
result and slighly smaller than the prediction of bosonized field 
theory. The relevance of 
these numerical results to real SP materials, such as CuGeO$_3$ and 
the recently discovered NaV$_2$O$_5$,\cite{navo} has to be determined 
experimentally.

The numerical procedures developed in this article could be applied 
to the study of several other properties of the incommensurate phase
of spin-Peierls systems,
such as the static magnetization as a function of the magnetic field 
(recently measured in
CuGeO$_3$ by Fagot-Revurat {\it et al.}\cite{fagot}) and the
order of the transition from the dimerized to the incommensurate
phases.\cite{loosd}


\begin{references}

\bibitem{haldane} F. D. M. Haldane, Phys. Rev. Lett. {\bf 50},
        1153 (1983).

\bibitem{ladders} For a recent review see {\it Physics Today},
        Search and Discovery, pg. 17 October 1996.

\bibitem{pytte} E. Pytte, Phys. Rev. B {\bf 10}, 2309 (1974).

\bibitem{hase} M. Hase, I. Terasaki and K. Uchinokura, Phys. Rev. 
        Lett. {\bf 70}, 3651 (1993).

\bibitem{regnault} L. P. Regnault {\it et al.}, Phys. Rev. B {\bf 53}, 
        5579 (1996).

\bibitem{lussier} J.-L. Lussier {\it et al.}, J. Phys. Condens.
        Matter {\bf 7}, 325 (1995).

\bibitem{cross2} M. C. Cross, Phys. Rev. B {\bf 20}, 4606 (1979).

\bibitem{hase2} M. Hase {\it et al.}, Phys. Rev. B {\bf 48}, 9616 
        (1993).

\bibitem{loosd} P. H. M. van Loosdrecht {\it et al.}, Phys. Rev. B
        {\bf 54}, 3730 (1996).

\bibitem{fujita} M. Fujita and K. Machida, J. Phys. Jpn {\bf 53}, 
        4395 (1984).

\bibitem{crossfisher} M.S. Cross and D.S. Fisher, Phys. Rev. B 
        {\bf 19}, 402 (1979).

\bibitem{nakano} T. Nakano and H. Fukuyama, J. Phys. Jpn {\bf 49}, 
        1679 (1980).

\bibitem{buzdin} A. I. Buzdin, M. L. Kulic, and V. V. Tugushev, Solid
        State Commun. {\bf 48}, 483 (1983).

\bibitem{kiry} V. Kiryukhin {\it et al.}, Phys. Rev. Lett. {\bf 76}, 
        4608 (1996); V. Kiryukhin {\it et al.}, Phys. Rev. B {\bf 54},
        7269 (1996).

\bibitem{fagot} Y. Fagot-Revurat {\it et al.}, Phys. Rev. Lett. 
        {\bf 77}, 1861 (1996).

\bibitem{affleck} I. Affleck,  {\it Fields, Strings and Critical 
        Phenomena}, edited by E. Br\'ezin and J.Zinn-Justin 
        (North-Holland, Amsterdam, 1990), pg. 563.

\bibitem{zang2} J. Zang, S. Chakravarty and A.R. Bishop, 
              cond-mat/9702185.

\bibitem{dobryriera} A. Dobry and J. Riera, (to be published).

\bibitem{rieradobry} J. Riera and A. Dobry, Phys. Rev. B {\bf 51}, 
        16098 (1995).

\bibitem{castilla} G. Castilla, S. Chakravarty and V.J. Emery, Phys. 
        Rev. Lett. {\bf 75}, 1823 (1995).

\bibitem{haas} S. Haas and E. Dagotto, Phys. Rev. B {\bf 52},
        14396 (1995).

\bibitem{rierakoval} J. Riera and  S. Koval, Phys. Rev. B {\bf 53}, 
        770 (1996).

\bibitem{poilblanc} D. Poilblanc {\it et al.}, Phys. Rev. B, to appear 
        (1997).

\bibitem{okamoto} K. Okamoto and K. Nomura, Phys. Lett. A
        {\bf 169}, 433 (1992).

\bibitem{spronken} G. Spronken, B. Fourcade, and Y. L\'epine, Phys.
        Rev. {\bf 33}, 1886 (1986), and references therein.

\bibitem{laukamp} Numerical calculations indicate that this relation
        still holds with an exponent $\nu$ close to 2/3, for 
        $\alpha > \alpha_c$, at least for not too small $\delta_0$ 
        and $\alpha < 1/2$; M. Laukamp and J. Riera, (to be 
        published).

\bibitem{fledder} A. Fledderjohann and C. Gros, cond-mat/9612013.

\bibitem{WLMC} J. E. Hirsch {\it et al.}, Phys. Rev. B {\bf 26},
         5033 (1982).

\bibitem{navo} M.~Isobe and Y.~Ueda, J.~Phys.~Soc.~Jpn. {\bf 65}, 1178 
        (1996); M.~Weiden, R.~Hauptmann, C.~Geibel, F.~Steglich,
        M.~Fischer, P.~Lemmens and G.~G\"untherodt, preprint 
        cond-mat/9703052.

\end{references}
\end{document}